\documentclass[a4paper,11pt]{article}
\usepackage{RR}

\usepackage{times}

\newcommand{\sq}{\hbox{\rlap{$\sqcap$}$\sqcup$}}
\newcommand{\qed}{\hspace*{\fill}\sq}

\newcommand{\myfontsize}{\fontsize{8}{10}\selectfont}

\usepackage[hang,small]{caption}
\usepackage{graphicx}
\usepackage{xspace}
\usepackage{subfig}
\usepackage{newclude}
\usepackage{amsmath,amsfonts,amssymb}
\usepackage{soul}
\usepackage{ulem}
\usepackage{adjustbox}
\usepackage[ruled,vlined]{algorithm2e}
\usepackage{comment}
\usepackage{multirow}
\usepackage[subject={Top1},author={Mathias}]{pdfcomment}
\usepackage{paralist}

\usepackage{pgf}
\usepackage{tikz}
\usetikzlibrary{calc,shapes,arrows,backgrounds,positioning}
\usetikzlibrary{decorations.pathmorphing}
\usetikzlibrary{decorations.pathreplacing}
\usetikzlibrary{decorations.shapes}
\usetikzlibrary{decorations.text}
\usetikzlibrary{decorations.markings}
\usetikzlibrary{decorations.fractals}
\usetikzlibrary{decorations.footprints}
\usepackage{gnuplot-lua-tikz}

\graphicspath{{figures/}}
\usepackage[version=3]{mhchem}

\input{macros}
\renewcommand{\ignoreconf}[1]{}

\RRdate{Mars 2013}
\RRauthor{
  Laura Grigori\thanks[inria]{INRIA Paris - Rocquencourt, B.P.
  105, F-78153 Le Chesnay Cedex, France}\thanks[upmc]{{UPMC Univ Paris 6, CNRS UMR 7598, Laboratoire Jacques-Louis Lions, F-75005, Paris, France}}
  \and 
    Mathias Jacquelin\thanksref{inria}\thanksref{upmc}
  \and 
    Amal Khabou\thanksref{inria}\thanksref{upmc}
}

\authorhead{L. Grigori, M. Jacquelin \& A. Khabou}
\RRtitle{Factorisations LU et QR multi-niveaux optimales en communication pour plates-formes hi\'erarchiques}
\RRetitle{Multilevel communication optimal LU and QR factorizations for hierarchical platforms}
\titlehead{Multilevel communication optimal LU and QR factorizations} 
\RRresume{
  Cette \'etude porte sur l'analyse des performances de deux algorithmes classiques de l'alg\`ebre lin\'eaire
  dense, les factorisations LU et QR, sur des plates-formes multi-niveaux hi\'erarchiques. Nous
  pr\'esentons tout d'abord un nouveau mod\`ele analytique appel\'e Hierarchical Cluster Platform
  (\HCP), encapsulant les caract\'eristiques de ce type de plates-formes. Plus pr\'ecis\'ement,
  l'emphase est mise sur ce qui se passe \`a chaque niveau de la hi\'erarchie. Nous \'etendons des bornes inf\'erieures
  sur les communications au mod\`ele \HCP. Nous introduisons ensuite deux algorithmes multi-niveaux
  adapt\'es \`a ces plates-formes pour les factorisations LU et QR, et analysons leurs performances.
  Nous pr\'esentons en outre un ensemble d'exp\'eriences num\'eriques ainsi que des pr\'edictions de
  performances illustrant la n\'ecessit\'e de tels algorithmes sur les plates-formes \`a grande
  \'echelle.
  }
\RRabstract{
This study focuses on the performance of two classical dense linear
algebra algorithms, the LU and the QR factorizations, on multilevel
hierarchical platforms. We first introduce a new model called Hierarchical
Cluster Platform (\HCP), encapsulating the characteristics of such platforms. The focus is
set on reducing the communication requirements of studied algorithms
at each level of the hierarchy. Lower bounds on communications are
therefore extended with respect to the \HCP model. We then introduce multilevel
LU and QR algorithms tailored for those platforms, and provide a detailed
performance analysis. We also provide a set of numerical experiments and
performance predictions demonstrating the need for such algorithms on large platforms. 
}

\RRmotcle{QR, LU, exascale, plates-formes hi\'erarchiques}
\RRkeyword{QR, LU, exascale, hierarchical platforms}
\RRprojet{ALPINES} 
\RCParis

\begin{document}

\RRNo{8270}
\makeRR

\section{Introduction}

Due to the ubiquity of multicore processors, solvers should be adapted
to better exploit the hierarchical structure of modern architectures,
where the tendency is towards multiple levels of parallelism. Thus with the
increasing complexity of nodes, it is important to exploit these many levels
of parallelism even within a single compute node. For that
reason, classical algorithms need to be revisited so as to fit
modern architectures that expose parallelism at different levels
in the hierarchy. We believe that such an approach is mandatory in order
to exploit upcoming hierarchical exascale computers at their full potential.

Studying the communication complexity of linear algebra operations and
designing algorithms that are able to minimize communication is a
topic that has received an important attention in the recent years.
The most advanced approach in this context assumes one level of
parallelism and takes into account the computation, the volume of
communication, and the number of messages exchanged along the critical
path of a parallel program.  In this framework, the main previous
theoretical result on communication complexity is a result derived by
Hong and Kung in the 80's providing lower bounds on the volume of
communication of dense matrix multiplication for sequential machines
\cite{hong1981io}.  This result has been extended to parallel machines
\cite{irony2004communication}, to dense LU and QR factorizations
(under certain assumptions)
\cite{demmel11:_commun_optim_paral_and_sequen}, and then to basically
all direct methods in linear algebra \cite{ballard2010minimizing}.
Given an algorithm that performs a certain number of floating point
operations, and given a size of the memory $M$, the lower
bounds on communication are obtained by using the Loomis-Whitney
inequality, as for example in \cite{irony2004communication,
  ballard2010minimizing}.  While theoretically important, these lower
bounds are derived with respect to a simple performance model that
supposes a memory hierarchy in the sequential case, and $P$ processors
without memory hierarchy in the parallel case.  Such a model is not
sufficient to encapsulate the features of modern architectures with
multilevel hierarchy.

On the practical side, several algorithms have been introduced
recently~\cite{tsqr-grid,SongLHD10,dague-la,Dongarra2013}. Most of them
propose to use multiple reduction trees depending on the hierarchy.
However, the focus is set on reducing the running time without explicitly taking
communication into consideration. In~\cite{Dongarra2013}, Dongarra et al. propose
a generic algorithm implementing several optimizations regarding pipelining of computation,
and allowing to select different elimination trees on platforms with two
levels of parallelism. They provide insights on choosing the appropriate tree,
a binary tree being for instance more suitable for a cluster with many cores, 
while a flat tree allows more locality and CPU efficiency.
However, no theoretical bounds nor cost analysis are provided in these studies
regarding communication.

In the first part of this paper we introduce a new model that we refer
to as the \HCP model. Provided that two supercomputers might have
different communication topologies and different compute nodes with
different memory hierarchies, a detailed performance model tailored
for one particular supercomputer is likely to not reflect the
architecture of another supercomputer.  Hence the goal of our
performance model is to capture the main characteristics that
influence the communication cost of peta- and exa- scale
supercomputers which are based on multiple levels of parallelism and
memory hierarchy.  We use the proposed \HCP model to extend the
existing lower bounds on communication for direct linear algebra, to
account for the hierarchical and heterogeneous nature of present-day
computers.  We determine what is the minimum amount of communication
that is necessary at every level in the hierarchy, in terms of both
number of messages and volume of communication.

In the second part of the paper we introduce two multilevel algorithms
for computing LU and QR factorizations (\MLCAQR and \MLCALU) that are
able to minimize the communication at each level of the hierarchy,
while performing a reasonable amount of extra computation. These recursive
algorithms rely on their corresponding 1-level algorithms (resp. \CAQR and \CALU)
as their base case. Indeed, these algorithms are known to attain the communication
lower bounds in terms of both bandwidth, and latency with respect to the simpler one level
performance model.

%

\section{Toward a realistic Hierarchical Cluster Platform model (\HCP)}
\label{mlcalumodel-sec.hcp}

\ignoreconf{
\begin{figure}[h!]
\centering
\begin{adjustbox}{width=.8\linewidth}
\includegraphics{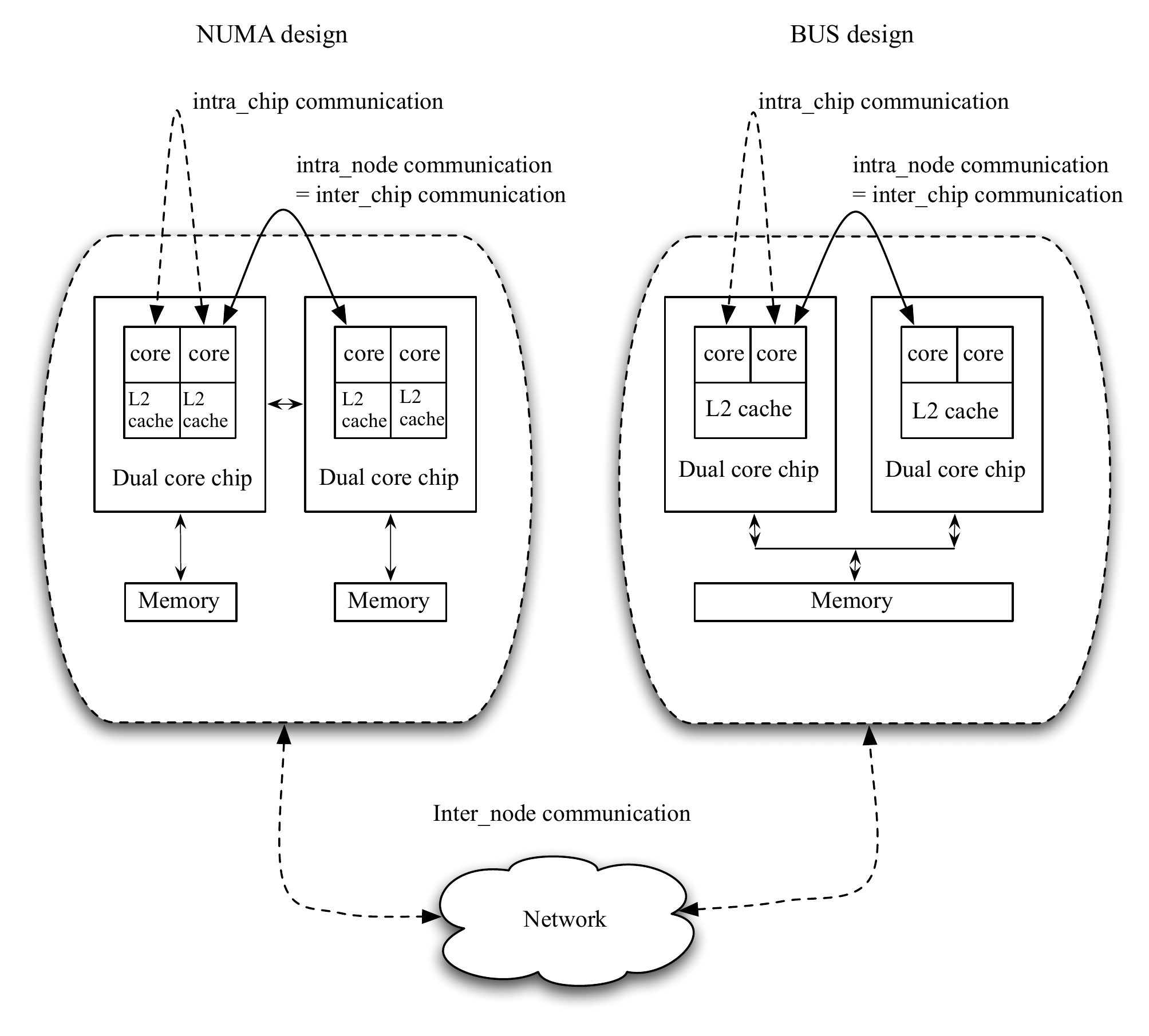}
\end{adjustbox}
\caption{\label{fig.hierarchical_cluster} Cluster of multicore processors.}
\end{figure}
}

The focus is set on hierarchical platforms implementing deeper and deeper
hierarchies. Typically, these platforms are composed of two kinds of
hierarchies: (1) a network hierarchy composed of interconnected
network nodes, which is stacked on top of a (2) computing nodes
hierarchy~\cite{HiHCoHPcappello}.  This compute hierarchy can be
composed for instance of a shared memory NUMA multicore platform.

\begin{figure}[htbp]
\centering
\begin{adjustbox}{width=.3\linewidth}
\begin{tikzpicture}

\node at (0,0) (lk) {Level $i$};
\node[draw=black,solid, minimum width = 1em, minimum height = 1em, right = 1cm of lk] (Pk0) {$CN_i$};
\node[draw=black,solid, minimum width = 1em, minimum height = 1em, right = 2cm of Pk0] (Pk1){$CN_i$};

\node[draw=black,dotted,fill=white, minimum width = 1em, minimum height = 1em] at ($(Pk0.east) + (1cm,1.5cm)$) (Bk) {$\buf{i+1}$};

\node[rectangle,draw=black,solid, minimum width = 1em, minimum height = 1em, text centered, above =1.5cm of Bk] (Pk2) {$CN_{i+1}$};

\node[rectangle,left = 2cm  of Pk2] (lk1) {Level $i+1$};

\node[draw=black,dotted, fill=white,minimum width = 1em, minimum height = 1em, below = 1cm of Pk0] (Bk0){$\buf{i}$};
\node[draw=black,dotted,fill=white, minimum width = 1em, minimum height = 1em, below = 1cm of Pk1] (Bk1){$\buf{i}$};

\draw [latex-latex,  thick] ($(Pk0.east)$) to ($(Pk1.west)$);

\draw [latex-latex,thick] (Pk0.south) to (Bk0.north);
\draw [latex-latex, thick] (Pk1.south) to (Bk1.north);

\draw [latex-latex,ultra thick] (Pk0.north) |- ($(Bk.south west) + (0.2cm,-0.5cm)$) -| ($(Bk.south west) + (0.2cm,0)$);
\draw [latex-latex,ultra thick] (Pk1.north) |- ($(Bk.south east) + (-0.2cm,-0.5cm)$) -| ($(Bk.south east) + (-0.2cm,0)$);

\draw [latex-latex,ultra thick] (Pk2.south) -- (Bk.north);

\draw (lk1.west |- Bk0.center) -- (Bk0.west);
\draw (Bk0.east) -- (Bk1.west);
\draw (Bk1.east) -- ($(Bk1.center) + (1cm,0)$);

\draw (lk1.west |- Bk.center)  -- (Bk.west);
\draw (Bk.east) -- ($(Bk1.center |- Bk.center) + (1cm,0)$);

\node[left] at ($0.5*(Bk) + 0.5*(Pk2) -(.2cm, 0) $) {$\beta_{i+1}$};

\node[left] at ($0.5*(Bk0) + 0.5*(Pk0) -(.2cm, 0) $) {$\beta_{i}$};
\node[right] at ($0.5*(Bk1) + 0.5*(Pk1) +(.2cm, 0) $) {$\beta_{i}$};

\node at ($0.5*(Pk0) + 0.5*(Pk1) -(0,.35cm) $) {$\beta_{i}$};
\node[left] at ($0.5*(Pk0) + 0.5*(Bk) +(.25cm, 0.25cm) $) {$\beta_{i+1}$};
\node[right] at ($0.5*(Pk1) + 0.5*(Bk) +(-.25cm, 0.25cm) $) {$\beta_{i+1}$};

\end{tikzpicture}
\end{adjustbox}
\caption{Components of a level $i$ in the \HCP model.\label{fig.level_model}}
\end{figure}
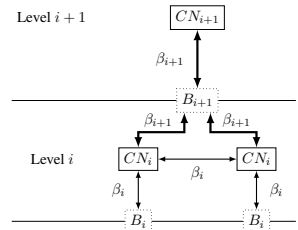

Our \HCP model considers such platforms with $l$ levels of parallelism, and
uses the following assumptions.  Level $1$ is the deepest level in the
hierarchy, where actual processing elements are located (for example
cores).
An intermediate level $i>1$ and its components, as depicted
in Figure~\ref{fig.level_model}, have the following characteristics.
A compute node of level $i+1$, denoted as $CN_{i+1}$ in the figure, is
formed by $\Proc{i}$ compute nodes of level $i$ (two nodes in
our example).  These $\Proc{i}$ compute nodes are organized along a
2D grid topology, that is $\Proc{i} = \Prow{i} \times \Pcol{i}$.  The
total number of compute nodes of the entire platform is $\Proc{} =
\prod_{i=1}^{l} \Proc{i}$, while the total number of compute nodes of level $i$
is denoted $\sProc{i} = \prod_{j=i}^{l} \Proc{j}$. 
We let $\mem{i}  = \mem{1} \cdot \prod_{j=1}^{i-1}{\Proc{j}}$ be
the aggregated memory size of a node of level $i>1$, 
 where \mem{1} denotes the memory size of
a processing element of level $1$. 
The network latency $\alpha_{i}$ and the inverse bandwidth $\beta_{i}$ apply throughout an entire level
$i$, however the higher in the hierarchy, the more important communication costs become.  We also
consider a network buffer $\buf{i}$ at each level of the hierarchy.  This allows to take into
account the possibility of message aggregation for network communication, and determine the number of
messages required to send a given amount of data, thus the latency cost associated with each
communication at every level of the hierarchy.  These notations will be used throughout the rest of
the paper.  In addition, we refer to the number of messages sent by a node of level $i$ as $S_i$,
and to the exchanged volume of data as $W_i$.  $\bar{S}_i$ is the latency cost at level $i$,
$\bar{S}_i = S_i\cdot \alpha_i$. Similarly, $\bar{W}_i$ is the bandwidth cost, $\bar{W}_i = W_i
\cdot \beta_i$.

We note that the model makes abstraction of the detailed architecture of a compute node or the
interconnection topology at a given level of the hierarchy.  Hence such an approach has its own
limitations, since the performance predicted by such a model might not be extremely accurate.
However, while keeping the model tractable, this model better reflects the actual nature of supercomputers
than the one level model assumed so far, and it will also allow us to understand the communication
bottlenecks of linear algebra operations.

\paragraph{Communicating under the \HCP model.}
We now describe how communication happens in the \HCP model, and how messages are routed between 
different levels of the network hierarchy. First, 
we assume that if a compute node of level $i$ communicates, all the nodes below
it participates. We denote as \textbf{\textit{counterparts}} of a compute node
of level $i$ all the nodes of level $i$ lying in remote compute nodes of level $i+1$
having the same local coordinates.
\ignoreconf{, leading to a total of $\prod_{j=1}^{i-1} \Proc{j}$  nodes of 
level 1 involved in that communication. Remember that $W_i$ is the amount of data that is sent by a node of level $i$. }
We therefore have the relation $W_i = W_{i+1} / \Proc{i}$.

As an example, let us detail a communication taking place between two compute nodes 
of a given level $i$. A total of $ \Proc{}/ \sProc{i} $ processing elements of level $1$ are involved in sending a global
volume of data $W_i$. Each of these elements has to send a chunk of data $W_1 = {W_i} \Proc{}/ \sProc{i}$. 
Since this amount of data has to fit in the memory of a processing element of level 1, we obviously have 
$\forall i, \mem{1} \geq W_1 = {W_i} \Proc{}/ \sProc{i} $. 
\ignoreconf{Here for the sake of simplicity, and to be able to derive 
general lower bounds valid for each level of the hierarchy, we assume the network buffer size
at level $1$ to be equal to the memory size of a compute unit of level $1$, $\buf{1} = \mem{1}$, which means that the first level of the network is fully-pipelined. Therefore, data is always sent using only one message 
at level $1$, by a compute unit at this level. The corresponding 
bandwidth cost is $W_1 \times \beta_1$.\\}
These blocks are transmitted to the level above in the hierarchy, i.e. to level $2$. 
A compute node of level $2$ has to send a volume of data $W_2 = \Proc{1} W_1$. 
Since the network buffer size at level $2$ is \buf{2}, this requires $({W_2}/{\buf{2}})$ messages.
The same holds for any level $k$ such that $1 < k \leq i$, where data is forwarded by 
sending$({W_k}/{\buf{k}})$ messages.  
We therefore have the following costs:

\[ \begin{array}{lcl}
  \bar{W}_k = \frac{W_i  \sProc{k}}{ \sProc{i} } \cdot \beta_k, &
  \quad \quad
  &\bar{S}_k = \frac{W_k}{\buf{k}} \cdot \alpha_{k} =\frac{W_i  \sProc{k}}{\buf{k} \sProc{i} } \cdot \alpha_k.
  \end{array} \]

\paragraph{Network types.} We assume three kinds of networks, depending on their respective buffer size:

\begin{compactenum}
\item \textit{fully-pipelined networks}, able to aggregate all incoming messages into a single message. 
This requires that $\buf{i} \geq \mem{i}$, since at most \mem{i} words of data can be sent.

The \textit{fully-pipelined network} case is ensured whenever  
$\buf{i} \geq \Proc{i-1} W_{i-1}$. Since \mem{i} is 
the size of the largest message sent at level $i$, we assume $\buf{i} = \mem{i}$ . We also assume 
that all levels below a fully-pipelined level are themselves fully-pipelined. Therefore, the constraint on the 
buffer size becomes $\buf{i}=\mem{i} = \Proc{i-1} \buf{i-1}$.
\item \textit{bufferized networks}, allowing for message aggregation up to a buffer size $\buf{i} <
\mem{i}$.
\item \textit{forward networks}, where messages coming from lower level are simply forwarded to
higher levels.

For a given level $i$, a \textit{forward network} requires that $\buf{i} = \buf{i-1}$. Indeed, when all the
sub-nodes from level $i-1$ send $S_{i-1}$ messages each, the number of forwarded messages is
$\bar{S}_{i} = \Proc{i-1} \bar{S}_{i-1}$.

\end{compactenum}


Based on the two extreme cases, we assume the buffer size \buf{i} to satisfy $\buf{i-1} \leq
\buf{i} \leq \Proc{i-1} \buf{i-1}$.

\paragraph{Lower bounds on communication.}
Lower bounds on communication have been generalized in~\cite{ballard2010minimizing} for
direct methods of linear algebra algorithms which can be expressed as three nested loops. 
We refine these lower bounds under our hierarchical model. For matrix product-like problems, 
at least one copy of the input matrix has to be stored in memory: a compute node of level $i$
thus needs a memory of $\mem{i} = \Omega( n^2/\sProc{i} )$.
Furthermore, the lower bound on latency depends on the buffer size \buf{i} of the considered
level $i$, where a volume $\bar{W}_i$ needs to be sent in messages of size
$\buf{i}$. Hence the lower bounds on communications at level $i$:

\centerline{
\begin{minipage}{.4\linewidth}
\begin{align}
    \bar{W_i} & \ignoreconf{\geq \Omega\left( \frac{n^3}{ \sProc{i} \sqrt{\mem{i}}} \cdot \beta_{i}
    \right)} = \Omega\left( \frac{n^2}{\sqrt{\sProc{i}}} \cdot \beta_{i} \right)
\end{align}
\end{minipage}
\begin{minipage}{.4\linewidth}
\begin{align}
    \bar{S_i} & \ignoreconf{\geq \Omega\left( \frac{n^3}{ \buf{i} \sProc{i} \sqrt{\mem{i}} } \cdot
    \alpha_{i}\right)} = \Omega\left( \frac{n^2}{\buf{i} \sqrt{\sProc{i}}}  \cdot
    \alpha_{i} \right)
\end{align}
\end{minipage}
}

\medskip

Note that, for simplicity, we expressed the bound on latency with respect to \buf{i} for all level
$i$. Since we consider $\buf{1} = \mem{1}$, the lower bound on latency for level $1$ can also be
expressed as $\bar{S}_1 = \Omega\left( \sqrt{\Proc{}} \right)$.

\section{Multilevel algorithms}

In this section, we introduce \MLCAQR and \MLCALU , two multilevel
algorithms for computing the QR and the LU factorizations of a dense
matrix $A$.  These multilevel algorithms heavily rely on their
relative 1-level communication optimal algorithms (\CAQR and \CALU),
and can be seen as a recursive version of these algorithms.  \MLCAQR
and \MLCALU recursive layout naturally allows for local elimination trees
tailored for hierarchical platforms, thus reducing the communication needs at
each level of the hierarchy.


\subsection{Multilevel QR factorization}

The QR factorization is a widely used algorithm. Example of use are numerous,
be it for orthogonalizing a set of vectors or for solving least squares problems.
It decomposes a matrix $A$ into two
matrices $Q$ and $R$ such that $A = QR$, where $Q$ is orthogonal and
$R$ is upper triangular. The decomposition is obtained by using
very stable transformations, such as Householder reflections.
We assume in the following that
the matrix $Q$ is not stored explicitly, but rather using the compact
$Y T Y^T$ representation \cite{schreiber1989storage}.



\MLCAQR, given in Algorithm~\ref{alg.mlcaqr}, is a
multilevel tree-based algorithm computing the QR factorization of a
matrix using Householder reflections. It is tailored for hierarchical
platforms. Moreover, as
\MLCAQR is a tree-based algorithm, these Householder reflectors are
stored in the lower triangular part of matrix $A$ using a tree
structure as in~\cite{demmel11:_commun_optim_paral_and_sequen}.

As \CAQR on platforms with one level of parallelism, \MLCAQR aims at reducing the amount of communication
required during the factorization, but at each level of parallelism of
a hierarchical platform.
At the topmost level of the hierarchy, \MLCAQR processes the entire input matrix $A$ panel by
panel. Each panel is first factored by a recursive call to \MLCAQR on the next lower level.
The Householder reflectors are then sent to remote compute nodes and the trailing matrix updated
using two recursive routines: \MLUPFACT and \MLUPELIM.

\begin{algorithm}[htbp]
\scalebox{0.7}{%
\begin{minipage}{1.2\linewidth}
\DontPrintSemicolon%
\If{$r=1$}{%
  Call $\CAQR(A,P)$\;%
}%
\Else{%
\For{$kk=1$ to $n/\bs{r}$}{%
  \For{Processor $p=1$ to $\Prow{r}$ in parallel}{%
    $h_p = (m-(kk-1)\bs{r})/\Prow{r}$\;%
    $panel = A(kk \cdot \bs{r}+(p-1)h_p:kk \cdot \bs{r}+p h_p,kk \cdot \bs{r}:(kk+1) \cdot \bs{r})$\;%
    Call $\MLCAQR(panel,h_p,\bs{r},r-1,p)$\;%
  }%
%
  \For{$j=1$ to $\log{\Prow{r}}$}{%
    ($p_{source}$,$p_{target}$) is the pair of processors with which this elimination is performed.\;%
    Send local \bs{r}-by-\bs{r} to the remote processor $p_{target}$ \;%
    Stack two \bs{r}-by-\bs{r} upper triangular matrices in $RR$ \;%
    Call $\MLCAQR(RR,2\bs{r},\bs{r},r-1,p_{source})$\;%
    Call $\MLCAQR(RR,2\bs{r},\bs{r},r-1,p_{target})$\;%
  }%
%
%
  \For{Processor $p=1$ to $\Prow{r}$ in parallel}{%
    Broadcast the sets of Householder vectors to every processor belonging to the same processor row\;%
    \For{Processor $pp=2$ to $\Pcol{r}$ on same row than $p$ in parallel}{%
      Call $\MLUPFACT(r-1,pp)$\;%
    }%
  }%
%
  \For{$j=1$ to $\log{\Prow{r}}$}{%
      ($p_{source}$,$p_{target}$) is the pair of processors with which this elimination was performed.\;%
      \For{Processor $pp=2$ to $\Pcol{r}$ on same row than $p_{source}$ in parallel}{%
        $pp_{target}$ is the remote processor on the same row than $p_{target}$ and in the same column than $pp$\;%
        $pp$ sends its local $C$ to $pp_{target}$\;%
        Call $\MLUPELIM(r-1,pp)$\;%
        Call $\MLUPELIM(r-1,pp_{target})$\;%
      }%
  }%
}%
}%
\end{minipage}
}

\caption{$\MLCAQR(A,m,n,r,P)$ \label{alg.mlcaqr}}

\end{algorithm}

More precisely, for each recursion level $r$, let \bs{r} be the block size, $m_s^{(r)}  = (m_s^{(r+1)}/\Prow{r+1} -
(s-1)\bs{r})$ be the panel row count at step $s$, and $n_s^{(r)} = (\bs{r+1} - s\bs{r})$ be the
number of columns in the trailing matrix. At the topmost level $l$, we have
$m_s^{(l)} = (m - (s-1)\bs{l})$ and $n_s^{(l)} = (n - s\bs{l})$.
\MLCAQR proceeds as follows: 

\begin{compactenum}
  \item The panel is factored by using a reduction operation, where
    \MLCAQR is the reduction operator.  With a binary tree, the computation becomes:
  \begin{compactenum}
    \item First \Prow{r} subsets of the panel, of size
          ${m_s^{(r)}}/{\Prow{r}}$-by-\bs{r}, are recursively factored with \MLCAQR (with a
          block size \bs{r-1} corresponding to the next level deeper in the hierarchy).
          At the deepest level of recursion, \MLCAQR calls the \CAQR algorithm.
    \item The resulting \bs{r}-by-\bs{r} $R$ factors are eliminated two-by-two
          using an elimination tree by multiple calls to \MLCAQR, requiring $\log{\Prow{r}}$
          steps along the critical path. 
  \end{compactenum}
  \item The current trailing matrix is then updated to reflect the factorization of the panel:
  \begin{compactenum}
    \item Updates corresponding to factorizations at the leaves of the tree are
    applied using the \MLUPFACT routine. This routine is called in parallel on \Prow{r} blocks rows
    of size ${m_s^{(r)}}/{\Prow{r}}$-by-$n_s^{(r)}$.
\MLUPFACT broadcasts \Prow{r} blocks of Householder's
reflectors of size ${m_s^{(r)}}/{\Prow{r}}$-by-\bs{r} from the column of
nodes holding current panel along rows of compute nodes. At the deepest level, the 
update corresponding to a leaf is applied as in \CAQR (see~\cite{lawn204}).

    \item Finally, the updates due to the eliminations of the intermediate $R$ factors are
    applied onto the trailing matrix using the \MLUPELIM procedure.    
Blocks of size $\bs{r}$-by-${n_s^{(r)}}/{\Pcol{r}}$
are exchanged within a pair of compute nodes. At the lowest level, a partial update is then computed locally
before being applied independently onto each processing elements, similarly to \CAQR
  \end{compactenum}
\end{compactenum}

\subsection{Multilevel LU factorizations}
\label{mlalg:mlcalu}

\ignoreconf{
\begin{algorithm}[htp]
\DontPrintSemicolon
\textbf{Input:} $m \times n$ matrix $A$, level of parallelism $l$ in the hierarchy, block size $\bs{l}$, 
number of nodes $\Proc{l} = \Prow{l} \times\Pcol{l}$ \;

\If{$l=1$}{
  Call $\CALU(A,m,n,1,P_1)$\;
}
\Else{

\For{$k=1$ to $n/\bs{l}$}{
   
   $m_p = (m-(k-1)\bs{l})/\Prow{l}$\;
   $n_p = (n-(k-1) \bs{l})/\Pcol{l}$\;

  /* factor leaves of the panel */ \;
  \For{Processor $p=1$ to $\Prow{l}$ \textbf{in parallel}}{
    
    $\textnormal{leaf} = A((k-1) \cdot \bs{l}+(p-1)m_p+1:(k-1) \cdot \bs{l}+p \cdot m_p,(k-1) \cdot \bs{l}+1:k \cdot \bs{l})$\;
    Call $\MLCALU(\textnormal{leaf}, m_p,\bs{l},l-1,\Proc{l-1})$\;
  }

  /* Reduction steps */ \;
  \For{$j=1$ to $\log{\Prow{l}}$}{
    Stack two \bs{l}-by-\bs{l} sets of candidate pivot rows in $B$\;
    Call $\MLCALU(B,2\bs{l},\bs{l},l-1,\Proc{l-1})$\;
  }

/* Compute block column of $L$ */   \;
 \For{Processor $p=1$ to $\Prow{l}$ \textbf{in parallel}}{
  Compute $L_{p,k}$ = $L(k \cdot \bs{l}+(p-1)m_p+1:k \cdot \bs{l}+p \cdot m_p, (k-1) \cdot \bs{l}+1:k \cdot \bs{l}) $\;
  /* ($L_{p,k} = L_{p,k} \cdot U_{k,k}^{-1}$) using multilevel algorithm with $P_{l-1}$ nodes at level $(l-1)$ */ \;
}

 /* Apply all row permutations */  \;
\For{Processor $p=1$ to $\Prow{l}$ \textbf{in parallel}}{
    Broadcast pivot information along the rows of the process grid \;
    /* all to all reduce operation using  $\Proc{l}$ processors of level $l$ */ \;
    Swap rows at left and right \;
}
  Broadcast right diagonal block of $L_{k,k}$ along rows of the process grid \;

/* Compute block row of $U$ */  \;
  \For{Processor $p=1$ to $\Pcol{l}$ \textbf{in parallel}}{
    Compute $U_{k,p}$= $U((k-1) \cdot \bs{l}+1:k \cdot \bs{l}, k \cdot \bs{l} + (p-1) n_p +1:  k \cdot \bs{l} + p \cdot n_p)$ \;
/* ($U_{k,p} = L_{k,k}^{-1} \cdot A_{k,p}^{-1}$) using multilevel algorithm with $P_{l-1}$ nodes at level $(l-1)$ */ \;
}
/* Update trailing matrix */ \;
 \For{Processor $p=1$ to $\Pcol{l}$ \textbf{in parallel}}{
 Broadcast $U_{k,p}$ along the columns of the process grid\;
}
 \For{Processor $p=1$ to $\Prow{l}$ \textbf{in parallel}}{
 Broadcast $L_{p,k}$ along the rows of the process grid\;
}
 \For{Processor $p=1$ to $\Proc{l}$ \textbf{in parallel}}{
$A(k \cdot \bs{l}+(p-1)m_p +1:k \cdot \bs{l}+p \cdot m_p,k \cdot \bs{l} + (p-1) n_p +1:  k \cdot \bs{l} + p \cdot n_p ) = $\;
\resizebox{0.95\linewidth}{!}{$A(k \cdot \bs{l}+(p-1)m_p +1:k \cdot \bs{l}+p \cdot m_p, k \cdot \bs{l} + (p-1) n_p +1:  k \cdot \bs{l} + p \cdot n_p) - L_{p,k}\cdot U_{k,p}$} \;
/* using multilevel Cannon with $P_{l}$ nodes at level $l$ */ \;
}
}
}
\caption{$\MLCALU(A,m,n,l,\Proc{l})$ \label{alg.mlcalu}}
\end{algorithm}
}

LU factorization is the cornerstone of many scientific computations,
and is the method of choice for solving most linear systems.  It
consists in decomposing a matrix $A$ into a lower triangular matrix
$L$ and an upper triangular matrix $U$ such that $PA=LU$, where $P$ is
a permutation matrix required for numerical stability reasons.  We present
two variants of a multilevel algorithm, \MLCALU, for computing the LU
factorization of a dense matrix.  \MLCALU is a recursive algorithm.
The first variant, 1D-\MLCALU is a uni-dimensional approach where the
entire panel is processed by a single recursive call. The second
variant, 2D-\MLCALU, processes a panel by multiple recursive calls on
sub-panels followed by a ``reduction'' phase similar to that of
\MLCAQR.  The base case of both recursive variants is
\CALU~\cite{Grigori:EECS-2010-29}, which uses tournament pivoting to
select pivots.

In the case of 1D-\MLCALU, the algorithm proceeds as follows at each
recursion level $r$.  (1) 1D-\MLCALU is recursively applied to
an entire panel of \bs{r} columns, only the number of compute nodes
along the columns varying.  (2) Once a panel is factored, a block of
rows of $U$ is computed by $\Pcol{r}\times \sProw{r}$ compute nodes of
level $r$. (3) The trailing matrix is finally updated after a
broadcast of the block column of $L$ along rows of the process grid
and the block row of $U$ along columns of the process grid. The matrix-matrix 
operations are performed using a multilevel matrix product algorithm, \MLCANNON,
which is based on the optimal Cannon algorithm. For more details, we
refer the interested reader to Algorithm~\ref{ch:mlcalumodel:ml-cannon} in
Appendix~\ref{app.mlcannon}.  At the
deepest level of recursion, panels of size $m\times \bs{1}$ are
factored by \CALU using $\Prow{}$ processing elements of level $1$.
Gaussian elimination with partial pivoting is first applied to blocks
of size $(m / \Prow{})$-by-$\bs{1}$, located at the leaves of the
reduction tree. These candidate pivot rows are then combined using
tournament pivoting, which involves communications at every level in
the hierarchy. Algorithm~\ref{1Dmlcalu} in Appendix~\ref{app.mlcalu} describes in 
details 1D \MLCALU In terms of numerical stability, 1D-\MLCALU is
equivalent to performing \CALU on a matrix of size $m\times n$, using
a block size $\bs{1}$ and a grid of processors $\Proc{} = \Prow{}
\times \Pcol{}$.

\ignore{The dashed black arrows correspond to 
the reduction operation, that is the application of GEPP to the blocks of the current panel. The black 
arrows represent the intra-node communication, which correspond to the communication between the compute 
nodes of level $1$. The red arrows represent a communication that involves two compute nodes belonging 
to two different nodes of level $2$. Finally the green arrows correspond to a communication between 
two compute units belonging to two different nodes of level $3$.
\begin{figure}[htbp]
\centering
\begin{minipage}[b]{.43\linewidth}
\centering
\begin{adjustbox}{scale=0.8}
\input{figures/tslu_reccalu.tikz}
\end{adjustbox}
\caption{TSLU on a hierarchical system with \\ three levels of parallelism. }
\label{tslu_reccalu}
\end{minipage}
\begin{minipage}[b]{.43\linewidth}
\centering
\begin{adjustbox}{scale=0.8}
\input{figures/caluprrp}
\end{adjustbox}
\caption{2D multilevel TSLU on a machine with two levels of parallelism. }
\label{fig:multilevel}
\end{minipage}
\end{figure}
}

The 2D-\MLCALU algorithm was first introduced in~\cite{DonfackGK12}
 and analyzed for two-levels platforms.  Here we extend the analysis of Algorithm~\ref{alg.mlcalu}, given in
Appendix~\ref{app.mlcalu}, to deeper platforms.  It proceeds as
follows: (1) the panel is recursively factored with 2D-\MLCALU with a
block size corresponding to the next level in the hierarchy.  Note
that at the deepest level of recursion, 2D-\MLCALU calls \CALU.  (2)
The selected sets of pivot candidates are merged two-by-two along the
reduction tree, where the reduction operator is 2D-\MLCALU. At the end
of the preprocessing step, the final set of pivot rows is selected and
each node working on the panel has the pivot information and the
diagonal block of $U$. (3) Then the computed permutation is applied to
the input matrix, the block column of $L$ and the block row of $U$ are
computed. (4) Finally, after the broadcast of $L$ and $U$ to
appropriate nodes, as in 1D-\MLCALU, the trailing matrix is updated
with \MLCANNON.



\section{Performance models}

In this section, we provide a cost analysis of both \MLCAQR and \MLCALU within the \HCP model.
We first analyse two recursive communication routines which are used by all multilevel
algorithms presented in this study, namely the point to point communication and the broadcast
operations. 

Point to point communication of a volume $D$ of data between two compute nodes of level $r$ involves compute nodes from level 1 to
level $r$. At every level, subnodes send their local data to their counterparts in the remote node
of level $r$.
The associated communication costs are:

\medskip
\centerline{
$
\begin{array}{ccc}
 \lBRecComm{W}{1\ldots r}{D} = \sum_{k=1}^{r} \frac{D \cdot \sProc{r}}{\sProc{k}}
\beta_{k}~, & \quad &
\lBRecComm{S}{1\ldots r}{D} = \alpha_{1} + \sum_{k=2}^{r} \frac{D \cdot
\sProc{r}}{\buf{k} \sProc{k}} \alpha_{k}~~.
\end{array}
$
}
\medskip

The broadcast operation between $\Pcol{r}$ compute nodes of level $r$ is very similar to point to
point communication, except that at every level, a node involved broadcasts its data to
$\Pcol{r}$ counterparts. A broadcast can thus be seen as $\log{\Pcol{r}}$ point to point communications.

\ignoreconf{
$\Pcol{r}$ nodes of level $r$ communicate in $\log{\Pcol{r}}$ steps at the topmost level, involving all their subnodes down to level 1.

Data is sent as follows: 
(1) at level $r$, a volume $D$ is sent along the row of \Prow{r} nodes of
level $r$.
(2) At each level $1 \leq k < r$, each compute node holding a block of the
panel of size $D \cdot \sProc{r}/ \sProc{k}$ broadcasts it to its counterparts in $\Pcol{k+1}$ compute nodes of level $k+1$.
The associated communication costs are given by:

\begin{minipage}[b]{0.35\linewidth}
\begin{equation*}
\scalebox{0.9}{
$
\lBRecBCast{W}{1\ldots r}{D} = \sum_{k=1}^{r-1}
\frac{ D \cdot \sProc{r} \log{\Pcol{r}} }{ \sProc{k} } \beta_{k}~~ ,
$
}
\end{equation*}
\end{minipage}
\begin{minipage}[b]{0.65\linewidth}
\begin{equation*}
\scalebox{0.9}{
$
\lBRecBCast{S}{1\ldots r}{D} {} =   \log{\Pcol{1}} \alpha_{1}
+ \sum_{k=2}^{r} \frac{ D \cdot \sProc{r} \log{\Pcol{r}}}{\buf{k} \sProc{k}} \alpha_{k} ~~ .
\label{eq.bcastlat}
$
}
\end{equation*}
\end{minipage}
}

\subsection{\MLCAQR}




We now review the cost of \MLCAQR in terms of computations as well as communications.  At each
recursion level $r$, parameters are adapted to a grid of \Prow{r}-by-\Pcol{r} compute nodes.  The
current panel is first factored by doing \Prow{r} parallel calls to \MLCAQR at the leaves of the
tree. Then, the resulting $R$ factors are eliminated through $\log{\Prow{r}}$ successive
factorizations of a $2\bs{r}$-by-$\bs{r}$ matrix formed by stacking up two upper triangular $R$ factors.
Once a panel is factored, the trailing matrix is updated. However, as the
Householder's reflectors are stored in a tree structure, like in \cite{lawn204}, the updates must be done by going through each level of the tree again.
These operations are recursively performed using \MLUPFACT for the leaves and \MLUPELIM for higher levels. 

\paragraph{Global recursive cost of \MLCAQR. }
We define the following contributions to the global cost of \MLCAQR:
\Bcaqr{m}{n}{b}{P} is the cost of factoring a matrix of size $m$-by-$n$ with \CAQR using $P$
processors and a block size $b$. \linebreak \Bmlcaqr{m}{n}{b}{P} is the cost of \MLCAQR on a $m$-by-$n$ 
matrix using $P$ processors and a block size $b$.
\Bupfact{m}{n}{b}{P} is the cost of updating the trailing matrix to reflect factorizations at
the leaves of the elimination trees. \Bupelim{m}{n}{b}{P} is the cost of applying updates
corresponding to higher levels in the trees.
In terms of communication, \MLUPFACT consists in broadcasting Householder reflectors
along process rows, while \MLUPELIM corresponds to $\log{\Prow{r}}$ point to point communications
of trailing matrix blocks between pairs of nodes.
Using these notations, the cost of \MLCAQR can be recursively expressed as:

\begin{equation}
\resizebox{0.65\linewidth}{!}{%
\begin{minipage}{1.2\linewidth}
\begin{alignat}{1}
\Bmlcaqr{m}{n}{\bs{r}}{\Proc{r}}  = & \left\{ %
                \begin{array}{ll}%
                ~~~
                \begin{aligned}%
                            & \sum_{s=1}^{n/\bs{r}} \left[ \Bmlcaqr{\frac{m -
                            (s-1)\bs{r}}{\Prow{r}}}{\bs{r}}{\bs{r-1}}{\Proc{r-1}} \right.\\%
                            & \quad \quad + \log{\Prow{r}} \cdot  \lBRecComm{T}{1\ldots r}{\frac{\bs{r}^2}{ 2}} \\
                            & \quad \quad \quad +  \log{\Prow{r}} \cdot  \Bmlcaqr{2\bs{r}}{\bs{r}}{\bs{r-1}}{\Proc{r-1}} \\%
                            & \quad \quad \quad \quad +  \Bupfact{\frac{m -(s-1)\bs{r}}{\Prow{r}}}{\frac{n - s\bs{r}}{\Pcol{r}}}{\bs{r-1}}{\Proc{r-1}} \\
                            & \left.\quad \quad \quad \quad \quad + \log{\Prow{r}} \cdot \Bupelim{2\bs{r}}{\frac{n-s\bs{r}}{\Pcol{r}}}{\bs{r-1}}{\Proc{r-1}} \right]
                \end{aligned} &  {\rm if} \; r>1 \\%
                \\%
                ~~~\Bcaqr{m}{n}{\bs{1}}{\Proc{1}} &  {\rm if} \; r=1 %
                \end{array}
                \right.%
\end{alignat}
\end{minipage}
}%
\end{equation}

\paragraph{Bounding cost of \MLCAQR. } \MLCAQR uses successive elimination trees
at each recursion depth $r$, each completed in $\log{\Prow{r}}$ steps. 
As successive trees from level $l$ down to level $r$ come from different
recursive calls, they are sequentialized. Thus, the total number of calls at a given recursion
depth $r$ can be upper-bounded by $N_{r} = 2^{l-r} \prod_{j=r}^{l} \log{\Prow{j}}$. 
The global cost of \MLCAQR can therefore be expressed in terms of number of calls at each level of
recursion, broken down between calls performed on leaves or higher level in the trees.
Details on this cost is given in Appendix~\ref{app.mlcaqr}.
Assuming that for each level $k$, we have $\Prow{k}=\Pcol{k}=\sqrt{\Proc{k}}$
, and block sizes chosen to make the additional costs lower order terms, that is
$\bs{k} = O(n/(\sqrt{\sProc{k}} \cdot \prod_{j=k}^{l}\log^2{\Proc{j}}))$,
the cost of \MLCAQR is:

\centerline{
\resizebox{.8\linewidth}{!}{
\begin{minipage}{1.2\linewidth}
\begin{alignat}{2}
&\sBmlcaqr{\bar{F}}{n}{n} && \leq \left[ \frac{4n^3}{\Proc{}} + O\left( \frac{l \cdot n^3}{\Proc{}
\prod_{j=1}^{l} \log{\Proc{j}}}\right) \right] \gamma \\
&\sBmlcaqr{\bar{W}}{n}{n} && \leq \left[ \frac{n^2}{\sqrt{\Proc{}}} \left( l\cdot \log{\Proc{1}} + \log{\Proc{l}} + 4l \cdot
\prod_{j=1}^{l} \log{\Proc{j}} \right) + O\left( \frac{l \cdot n^2}{\sqrt{\Proc{}}
\log{\Proc{l}}}\right) \right] \beta_1 \notag\\
& && \quad + \sum_{k=2}^{l-1} \left[\frac{(l-k) \cdot n^2}{\sqrt(\sProc{k})}\left( 1+ \frac{2
\prod_{j=k}^{l}\log{\Proc{j}}}{\sqrt{\Proc{l}}}\right) +O\left(\frac{(l-k) \cdot
n^2}{\sqrt{\sProc{k}} \log{\Proc{l}}}\right)  \right] \beta_k \\ 
& && \quad \quad + \left[ \frac{n^2}{\sqrt{\sProc{l}}} \cdot \log{\Proc{l}} + O\left( \frac{n^2}{\sqrt{\sProc{l}}
\log{\Proc{l}}}\right) \right] \beta_l
\notag \\
%
%
&\sBmlcaqr{\bar{S}}{n}{n} &&\leq
\left[ l \cdot \sqrt{\Proc{}}\cdot \prod_{j=1}^{l}\log^3{\Proc{j}} + O\left( \sqrt{\Proc{}}\cdot
\prod_{j=1}^{l}\log^2{\Proc{j}} \right) \right] \alpha_1 \notag \\
& && \quad + \sum_{k=2}^{l-1} \left[ \frac{n^2}{\buf{k}\sqrt{\sProc{k}}}\cdot (l-k)\log{\Proc{k}}
+O\left( \frac{(l-k) \cdot n^2}{\buf{k} \sqrt{\sProc{k}} \log{\Proc{l}}} \right) \right]
\alpha_k  \\
& && \quad \quad + \left[ \frac{n^2}{\buf{l}\sqrt{\Proc{l}}} \cdot \log{\Proc{l}} +
\frac{n^2}{\buf{l}\sqrt{\Proc{l}}\prod_{2}^{l-1}\sqrt{\Proc{j}} } \cdot \log{\Proc{l}} + O\left(
\frac{n^2}{\buf{l} \sqrt{\Proc{l} \log{\Proc{l}}}} \right) \right]
\alpha_l \notag
\end{alignat}
\end{minipage}
}
}
\medskip

Altogether, though the recursive nature of \MLCAQR leads to at least three times more computations than the
optimal algorithm, which is similar to other recursive approaches~\cite{frens-wise}, it allows to reach
the lower bound at all levels of the hierarchy up to polylogarithmic factors.
Indeed, chosing appropriate block sizes makes most of the extra computational costs lower order
while maintaining the optimality in terms of communications.

\subsection{2D multilevel CALU}
\label{mlcalumodel-sec.multi}

In this section we only detail the cost of 2D-\MLCALU with respect to the \HCP model. Thus, for simplicity, we refer 
to it as \MLCALU throughout the rest of the paper. We note that we use the same reasoning 
as \MLCAQR to derive the recursive cost of \MLCALU, and the same approach and approximations to estimate its total cost.
Thus for a square $n$-by-$n$ matrix and using $l$ levels of recursion ($l\geq 2$) the cost of \MLCALU is:

%
%
%
%
%
%

\centerline{
\resizebox{.8\linewidth}{!}{
\begin{minipage}{1.2\linewidth}
\begin{alignat}{2}
&\sBmlcalu{\bar{F}}{n}{n} && \leq \left[ \frac{2n^3}{3P}  + \frac{n^3}{P \log^2P_l} + \frac{n^3}{P}
\Big(\frac{3}{8}\Big)^{l-2} \Big( \frac{5}{16} l - \frac{53}{128} \Big) + O\left(
\frac{n^2}{\sqrt{\Proc{}}}\right) \right] \gamma 
\label{eq.mlcalu_flops}\\
&\sBmlcalu{\bar{W}}{n}{n} && \leq \left[ \frac{n^2}{ 2\sqrt{\Proc{}}}  \log\Proc{1} \prod_{j=2}^l
(1+ \frac{1}{2}\log\Proc{j}) \right]  \beta_1 \label{eq.mlcalu_bw}\\
& && \quad + \sum_{k=1}^{l} \Big[\frac{n^2}{  \sqrt{\Proc{k}^*}} \Big( \frac{8}{3} \log^2P_l
(1+\frac{l-k}{\sqrt{P_k}}) + \frac{(l-2)}{8} (1+ \frac{l}{4}) \prod_{j=3}^l (1+
\frac{1}{2}\log\Proc{j}) \Big)  \Big]\beta_k. \notag\\
&\sBmlcalu{\bar{S}}{n}{n} &&\leq \left[ \frac{n^2}{ 2\sqrt{\Proc{}}}  \log\Proc{1} \prod_{j=2}^l (1+
\frac{1}{2}\log\Proc{j}) \right]  \alpha_1  \label{eq.mlcalu_lat}\\
& && \quad + \sum_{k=1}^{l} \Big[\frac{n^2}{ \buf{k}\sqrt{\Proc{k}^*}} \Big( \frac{8}{3} \log^2P_l
(1+\frac{l-k}{\sqrt{P_k}}) + \frac{(l-2)}{8} (1+ \frac{l}{4}) \prod_{j=3}^l (1+
\frac{1}{2}\log\Proc{j}) \Big)  \Big] \alpha_k \notag
\end{alignat}
\end{minipage}
}
}
\medskip

Equation~\ref{eq.mlcalu_flops} shows that \MLCALU performs more floating-point operations than
\CALU. This is because of the recursive 
calls during the panel factorization. 
Note that certain assumptions should be done regarding the hierarchical structure of the computational system in order to 
keep the extra flops as a low order term, and therefore asymptotically reach the lower bounds on computation. 

In terms of communications, we can conclude that \MLCALU attains the lower bounds derived in section~\ref{mlcalumodel-sec.hcp} modulo a factor that 
depends on $l^2 \prod_{j=2}^l \log\Proc{j}$ at each level $k$ of hierarchy. Thus it reduces the 
communication cost at each level of a hierarchical system. Note that in practice the number of levels is 
going to remain small, while the number of processors will be large.
We note that we do not give the detailed cost of 1D-\MLCALU here. However we would like to point that it 
attains the lower bounds derived under the \HCP model in terms of bandwidth 
at each level of parallelism. However in terms of latency the lower bound is only met at the deepest 
level of parallelism. 

\section{Experimental results}
\subsection{Numerical stability of \MLCALU}
\label{subsec:multilevel_stability}
\label{sec:ml_experiments}
Since \MLCALU is based on recursive calls, its stability can be different from that of CALU.  
 Our experiments show that up to three 
levels of parallelism \MLCALU exhibits a good stability, however
further investigation is required if more than three levels of parallelism 
are used. 
We study both the stability of the LU decomposition and of the linear solver, 
in terms of growth factor and three different backward errors: the normwise 
backward error, the componentwise backward error, and the relative error 
$\| PA - LU \| / \|A \|$.

\begin{figure}[htbp]
\centering
    \includegraphics[width=.9\linewidth]{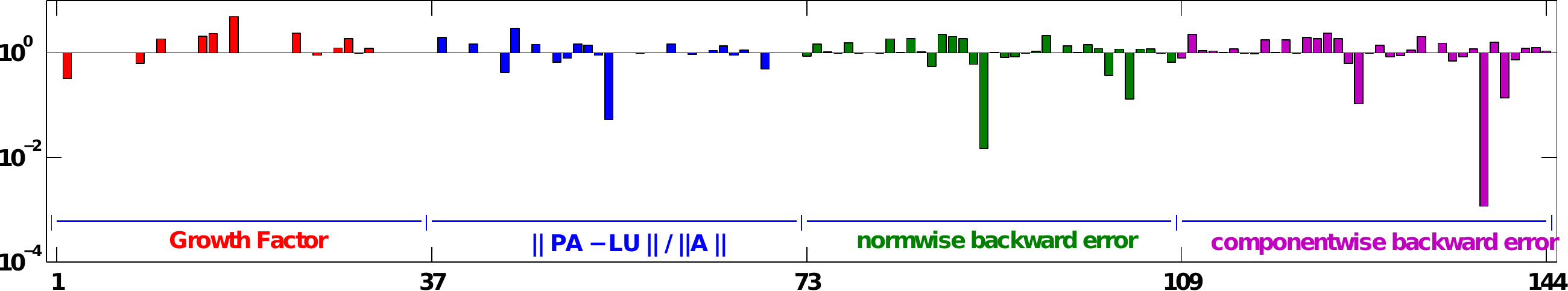}
    \caption{Ratios of 3-level \CALU's growth factor and backward errors to GEPP's.}
    \label{fig:error3l}
    \vspace{-0.5cm}
\end{figure}

Figure \ref{fig:error3l} displays the values of the ratios of 3-level CALU's
growth factor and backward errors to those of GEPP for 36 special matrices~\cite{Grigori:EECS-2010-29}.
The tested matrices are of size 8192, using the following parameters: $\Prow{3}=16$, $b_3=64$, 
$\Prow{2}=4$, $b_2=32$, $\Prow{1}=4$, and $b_1=8$.
We can see that nearly all ratios are between $0.002$ and $2.4$ for all tested
matrices. For the growth factors, the ratio is of order $1$ in $69 \%$ of the cases. For the
relative errors, the ratio is of order $1$ in $47 \%$ of the cases. 

We note that in most cases, ML-CALU uses tournament pivoting to select pivots at each level of the 
recursion, which does not ensure that the element of maximum magnitude in the column is used as pivot, 
neither at each level of the hierarchy, nor at each step of the LU factorization, that is
globally for the panel. For that reason we consider a threshold $\tau_k$, defined as the quotient of the pivot 
used at step $k$ divided by the maximum value in column $k$. 
We observe that in practice 
the pivots used by recursive tournament pivoting are close to the elements of 
maximum magnitude in the respective columns for both binary tree based 2-level CALU
and binary tree based 3-level CALU. For example, for binary tree based 3-level CALU, the selected pivot rows are equal to the elements 
of maximum magnitude in $63\% $  of the cases, and for the rest of the cases the minimum 
threshold $\tau_{\min}$ is larger than $0.30$. 

\subsection{Performance predictions}
\label{mlcalumodel-sec.ped}

In this section, we present performance predictions on a sample exascale platform. 
Current petascale platforms already display a hierarchical nature which strongly impacts
the performance of parallel applications. Exascale will dramatically amplify this trend. 
We plan here to provide an insight on what could be observed on such platforms.

We model the platform with respect to the \HCP model, and use it to estimate the running
times of our algorithms. As exascale platforms are not available yet, we base our sample
exascale platform on the characteristics of NERSC Hopper~\cite{hopper,extedhopper},
a petascale platform planned to reach the Exascale around year 2018.
It is composed of \textit{Compute Nodes}, each with two hexacore AMD Opteron Magny-cours
2.1GHz processors offering a peak performance of 8.4 GFlop/s, with 32 GB of memory.
Nodes are connected in pairs to \textit{Gemini ASICs}, which are 
interconnected through the \textit{Gemini network}~\cite{SURVEY1,SURVEY2}.
Detailed parameters of the Hopper platform are presented in Table~\ref{tab.exa_hopper}.

The target exascale platform is obtained by increasing the number of
nodes at all 3 levels, leading to a total of $1M$ nodes. The amount
 of memory per processing element is kept constant at 1.3 GB, while latencies and bandwidths are
 derived using an average $15\%$ decrease per year for the latency
 and a $26\%$ increase for the bandwidth~\cite{SURVEY2,SURVEY1}.
 These parameters are detailed in Table~\ref{tab.exa_hopper}.

\begin{table}[htbp]
\centering
\resizebox{\linewidth}{!}{
 \begin{tabular}{| c | c | c  | c | c || c | c | c | c |}
\hline
& \multicolumn{4}{c ||}{NERSC Hopper} & \multicolumn{4}{| c |}{Exascale platform}\\
\hline
Level & Type & \#  & Bandwidth & Latency & Type & \#  & Bandwidth & Latency \\
\hline
1 & 2x 6-cores Opterons & 12  & 19.8 GB/s & $1 \times 10^{-9}$s & Multi-cores
& 1024 &  300 GB/s & $1 \times 10^{-10}$s\\
2 & Hopper nodes & 2 & 10.4 GB/s & $1 \times 10^{-6}$s & Nodes  & 32 & 150 GB/s & $1 \times 10^{-7}$s \\
3 & Gemini ASICS & 9350  & 3.5 GB/s &  $1.5 \times 10^{-6}$s & Interconnects  & 32768 & 50 GB/s &  $1.5 \times 10^{-7}$s  \\
\hline
\end{tabular}
}
\caption{Characteristics of NERSC Hopper and sample exascale platform. \label{tab.exa_hopper} }
\end{table}

Moreover, in order to assess the performance of multilevel algorithms, costs of state-of-the-art
1-level communication avoiding algorithms need to be expressed in the \HCP model. To this end, we assume (1) each
communication to go through the entire hierarchy: two communicating nodes thus belong to two distant
nodes of level $l$, hence a bandwidth $\beta_l$. (2) Bandwidth is shared among parallel communications.

We evaluate the performance of the \MLCAQR and
\MLCALU algorithms as well as their corresponding 1-level routines on a matrix of size $n \times n$,
distributed over a square 2D grid of $P_k$ processors at each level 
$k$ of the hierarchy, $\Proc{k}= \sqrt{\Proc{k}}\times \sqrt{\Proc{k}}$.
In the following, we assume all levels to be \textit{fully-pipelined}. Similar results are obtained regarding
\textit{forward} hierarchies, which is explained by the fact that realistic test cases are not latency bounded.

\begin{figure}[htbp]
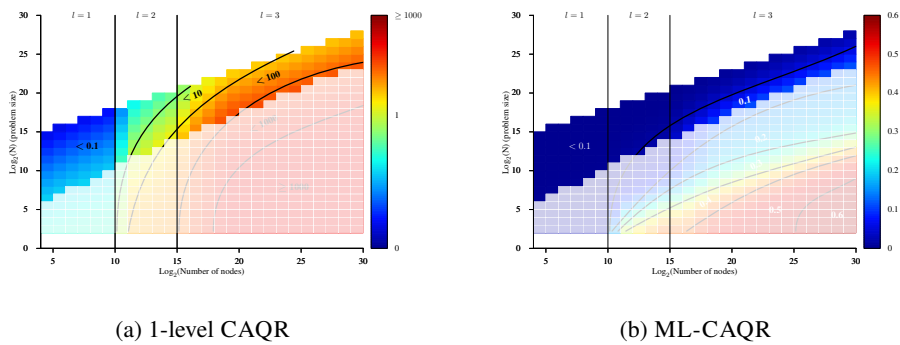

\centering
\subfloat[1-level \CAQR]{
\label{fig.comm_comp_caqr}
\begin{adjustbox}{width=.48\linewidth}
\input{figures/caqr_comm_comp.tikz}
\end{adjustbox}
}~~
\subfloat[\MLCAQR]{
\label{fig.comm_comp_mlcaqr}
\begin{adjustbox}{width=.48\linewidth}
\input{figures/mlcaqr_comm_comp.tikz}
\end{adjustbox}
}

\caption{Prediction of communication to computation ratio on an exascale platform.}
\label{fig.comm_comp}
\end{figure}

\paragraph{Performance predictions of \MLCAQR}
The larger the platform is, the more expensive the communications become. This trend can be
illustrated by observing the communication to computation ratio, or \textit{CCR} of an algorithm.
On Figure~\ref{fig.comm_comp}, we plot the \textit{CCR} of both \CAQR and \MLCAQR on the exascale
platform. The shaded areas correspond to unrealistic cases where there are more processing elements
than matrix elements. As the number of processing elements increases, cost of \CAQR (on
Figure~\ref{fig.comm_comp_caqr}) is dominated by communication.
Our multilevel approach alleviates this trend, and \MLCAQR (on Figure~\ref{fig.comm_comp_mlcaqr})
allows to maintain a good computational density, especially when the number of levels involved is
large. Note that for $l=1$, \MLCAQR and \CAQR are equivalent.

\begin{figure}[htbp]
\centering
\begin{minipage}{0.48\linewidth}
\begin{adjustbox}{width=\linewidth}
\input{figures/mlcaqr_speedup.tikz}
\end{adjustbox}
\caption{Speedup of \MLCAQR vs. \CAQR \label{fig.speedup_mlcaqr}}
\end{minipage}~~
\begin{minipage}{0.48\linewidth}
\begin{adjustbox}{width=\linewidth}
\input{figures/mlcalu_speedup.tikz}
\end{adjustbox}
\caption{Speedup of \MLCALU vs. \CALU \label{fig.speedup_mlcalu}}
\end{minipage}

\end{figure}

However, as \MLCAQR perform more computations than \CAQR, we compare the expected running times of
both algorithms. Here by running time, we denote the sum between computational and communication costs.
We thus assume no overlap between computations and communications. 
The ratio of the \MLCAQR running time over \CAQR is depicted on Figure~\ref{fig.speedup_mlcaqr}.
\MLCAQR clearly outperforms \CAQR when using the entire platform, despite
its higher computational costs. As a matter of a fact in this domain, the running time is
dominated by the bandwidth cost, and \MLCAQR significantly reduces it at all levels.

\paragraph{Performance predictions of \MLCALU} The same 
observations can be made on the \textit{CCR} of \CALU and
\MLCALU, we will therefore not present the detail here. Regarding the
running times ratio, depicted on Figure~\ref{fig.speedup_mlcalu}, we can also
conclude that \MLCALU is able to keep communication costs significantly
lower than \CALU, leading to significant performance improvements.

Altogether, our performance predictions validate our multilevel approach for
large scale hierarchical platforms that will arise with the Exascale. Indeed,
by taking communications into consideration at all levels, \MLCAQR and \MLCALU
deliver a high level of performance, even at scales where performance is hindered
by communication costs with 1-level communication avoiding algorithms.

\section{Conclusion}

In this paper we have introduced two algorithms, ML-CALU and ML-CAQR,
that minimize communication over multiple levels of parallelism at the
cost of performing redundant computation.  The complexity analysis is
performed by using HCP, a model that takes into account the cost of
communication at each level of a hierarchical platform.  The multilevel
QR algorithm has similar stability properties to classic algorithms.
Two variants of the multilevel LU factorization are discussed.  A
first variant, based on a uni-dimensional recursive approach, has the
same stability as CALU. However, while it minimizes bandwidth over
multiple levels of parallelism, it allows to minimize latency only
over one level of parallelism.  The second variant which uses a
two-dimensional recursive approach, is shown to be stable in practice,
and reduces both bandwidth and latency over multiple levels of
parallelism.

Our performance predictions on a model of an exascale platform show
that for strong scaling, the multilevel algorithms lead to important
speedups compared to the algorithms which minimize communication over
only one level of parallelism.  In most of the cases, minimizing
bandwidth is the key factor for improving scalability, and hence the
1D ML-CALU is an appropriate choice for both ensuring numerical
stability and being efficient in practice.



\clearpage
\bibliographystyle{plain}
\bibliography{multilevel}


\pagebreak \small

\begin{appendix}

\section{Appendix: \MLCANNON \label{app.mlcannon}}

To perform matrix multiplication $C = C + A \cdot B$, we can use Cannon's algorithm. It is known that (see for example~\cite{ballard2010minimizing}) 
assuming minimal memory usage, Cannon's algorithm is asymptotically optimal in terms of both bandwidth, and latency, that is 
the volume of data and the number of messages send by each node are asymptotically optimal. In the following we present a recursive Cannon algorithm over a hierarchy of nodes. 
We refer to this algorithm as ML-CANNON, and we consider $l$ levels of parallelism. Algorithm~\ref{ch:mlcalumodel:ml-cannon} 
presents the communication details of \MLCANNON.

\begin{algorithm}[h!]
\myfontsize{
\DontPrintSemicolon

\textbf{Input}: three square matrices $C$, $A$, and $B$,  $k$ the number of recursion level, $P_k$ the number of processors at level $k$\;
\textbf{Output}: $C = C + A B$\;
\If{$k==1$}{
$ CANNON (C, A, B, P_1)$\;
}
\Else{
\For{ $i=1$ to $\sqrt{P_k}$ \textbf{\textbf{in parallel}} }{
  
 left-circular-shift row $i$ of A by $i$\;
  
}
\For{ $i=1$ to $\sqrt{P_k}$ \textbf{in parallel}}{
  
up-circular-shift column $i$ of B by $i$\; 
}
\For{$h=1$ to $\sqrt{P_k}$ (sequential)}{
\For{ $i=1$ to $\sqrt{P_k}$  and $j=1$ to $\sqrt{P_k}$ \textbf{in parallel}}{

$ML-CANNON(C(i,j), A(i,j), B(i,j), P_{k-1}, k-1)$\;

 left-circular-shift row $i$ of A by $1$\;
 up-circular-shift column $i$ of B by $1$\;

}
}
}
\caption{\MLCANNON$(C, A, B, P_k, k)$}
\label{ch:mlcalumodel:ml-cannon}
}
\end{algorithm}

\section{Appendix: 1D-\MLCALU and 2D-\MLCALU \label{app.mlcalu}}
\paragraph{1D multilevel CALU.}

1D \MLCALU is described in Algorithm~\ref{1Dmlcalu}. 
It is applied to a matrix $A$ of size $m \times n$, using $l$ levels of recursion and a 
total number of $\Proc{} = \prod_{i=1}^l \Proc{i}$ compute nodes of level $1$ at the topmost 
level of the recursion. These compute nodes are organized as a two-dimensional grid at each level, $\Proc{i} =\Prow{i} \times\Pcol{i}$.

\begin{algorithm}[h!]
\myfontsize{
\DontPrintSemicolon

\textbf{Input:} $m \times n$ matrix $A$, the recursion level $l$, block size $b_l$, 
the total number of compute units $\Proc{} = \prod_{i=1}^l \Proc{i}=\Prow{} \times\Pcol{}$ \;
\If{ $l == 1$} {
$[\Pi_1, L_1, U_1] = \CALU(A,b_1,\Prow{}\times \Pcol{1})$ \;
}
\Else{
$M = m/b_l, N = n/b_l$ \;
\For{$K = 1$ to $N $}{

$[\Pi_{KK}, L_{K:M, K}, U_{KK}] = Recursive-\CALU(A_{K:M,K} ,l-1,b_{l-1},\Prow{}\times \prod_{i=1}^{l-1}\Pcol{i}  )$\;
/* Apply permutation and compute block row of $U$ */ \;
$A_{K:M, :} = \Pi_{KK} A_{K:M, :}$\;
\For{ \textbf{each} compute unit at level $l$ owning a block
  $A_{K,J}$, $J = K+1$ to $N$  \textbf{in parallel}} {
$U_{K,J} = L_{KK}^{-1} A_{K,J} $\;
/* call multilevel dtrsm using $\Pcol{l}\times \prod_{i=1}^{l-1}\Proc{i} $ processing node of level $1$ */\;
}
/* Update the trailing submatrix */\;
\For{ \textbf{each} compute unit at level $l$ owning a block $A_{I,J}$ of the trailing submatrix, $I, J = K+1$ to $M,N$  \textbf{in parallel}}{
$A_{I,J}$ = $A_{I,J} - L_{I,K} U_{K,J}$\;
/* call multilevel dgemm using $\Prow{}\times \prod_{i=1}^{l}\Pcol{i}$ processing node of level $1$ */\;
}
}
}

\caption{1D multilevel communication avoiding LU factorization}
\label{1Dmlcalu}
}

\end{algorithm}

\clearpage
\paragraph{2D multilevel CALU.}
The 2D \MLCALU algorithm, given in Algorithm~\ref{alg.mlcalu}, details the 
communication performed during the factorization.

\begin{algorithm}[h!]
\myfontsize{
\DontPrintSemicolon
\textbf{Input:} $m \times n$ matrix $A$, level of parallelism $l$ in the hierarchy, block size $\bs{l}$, 
number of nodes $\Proc{l} = \Prow{l} \times\Pcol{l}$ \;

\If{$l=1$}{
  Call $\CALU(A,m,n,1,P_1)$\;
}
\Else{

\For{$k=1$ to $n/\bs{l}$}{
   
   $m_p = (m-(k-1)\bs{l})/\Prow{l}$\;
   $n_p = (n-(k-1) \bs{l})/\Pcol{l}$\;

  /* factor leaves of the panel */ \;
  \For{Processor $p=1$ to $\Prow{l}$ \textbf{in parallel}}{
    
    $\textnormal{leaf} = A((k-1) \cdot \bs{l}+(p-1)m_p+1:(k-1) \cdot \bs{l}+p \cdot m_p,(k-1) \cdot \bs{l}+1:k \cdot \bs{l})$\;
    Call $\MLCALU(\textnormal{leaf}, m_p,\bs{l},l-1,\Proc{l-1})$\;
  }

  /* Reduction steps */ \;
  \For{$j=1$ to $\log{\Prow{l}}$}{
    Stack two \bs{l}-by-\bs{l} sets of candidate pivot rows in $B$\;
    Call $\MLCALU(B,2\bs{l},\bs{l},l-1,\Proc{l-1})$\;
  }

/* Compute block column of $L$ */   \;
 \For{Processor $p=1$ to $\Prow{l}$ \textbf{in parallel}}{
  Compute $L_{p,k}$ = $L(k \cdot \bs{l}+(p-1)m_p+1:k \cdot \bs{l}+p \cdot m_p, (k-1) \cdot \bs{l}+1:k \cdot \bs{l}) $\;
  /* ($L_{p,k} = L_{p,k} \cdot U_{k,k}^{-1}$) using multilevel algorithm with $P_{l-1}$ nodes at level $(l-1)$ */ \;
}

 /* Apply all row permutations */  \;
\For{Processor $p=1$ to $\Prow{l}$ \textbf{in parallel}}{
    Broadcast pivot information along the rows of the process grid \;
    /* all to all reduce operation using  $\Proc{l}$ processors of level $l$ */ \;
    Swap rows at left and right \;
}
  Broadcast right diagonal block of $L_{k,k}$ along rows of the process grid \;

/* Compute block row of $U$ */  \;
  \For{Processor $p=1$ to $\Pcol{l}$ \textbf{in parallel}}{
    Compute $U_{k,p}$= $U((k-1) \cdot \bs{l}+1:k \cdot \bs{l}, k \cdot \bs{l} + (p-1) n_p +1:  k \cdot \bs{l} + p \cdot n_p)$ \;
/* ($U_{k,p} = L_{k,k}^{-1} \cdot A_{k,p}^{-1}$) using multilevel algorithm with $P_{l-1}$ nodes at level $(l-1)$ */ \;
}
/* Update trailing matrix */ \;
 \For{Processor $p=1$ to $\Pcol{l}$ \textbf{in parallel}}{
 Broadcast $U_{k,p}$ along the columns of the process grid\;
}
 \For{Processor $p=1$ to $\Prow{l}$ \textbf{in parallel}}{
 Broadcast $L_{p,k}$ along the rows of the process grid\;
}
 \For{Processor $p=1$ to $\Proc{l}$ \textbf{in parallel}}{
$A(k \cdot \bs{l}+(p-1)m_p +1:k \cdot \bs{l}+p \cdot m_p,k \cdot \bs{l} + (p-1) n_p +1:  k \cdot \bs{l} + p \cdot n_p ) = $\;
$A(k \cdot \bs{l}+(p-1)m_p +1:k \cdot \bs{l}+p \cdot m_p, k \cdot \bs{l} + (p-1) n_p +1:  k \cdot \bs{l} + p \cdot n_p) - L_{p,k}\cdot U_{k,p}$ \;
/* using multilevel Cannon with $P_{l}$ nodes at level $l$ */ \;
}
}
}
\caption{2D multilevel communication avoiding LU factorization \label{alg.mlcalu}}
}
\end{algorithm}

\clearpage
\section{Appendix: detailed upper bound on the cost of \MLCAQR \label{app.mlcaqr}}

The global recursive cost of \MLCAQR can be upper bounded by:

\begin{equation*}
\begin{array}{l}
\sBmlcaqr{T}{m}{n} \leq \\ \\
\quad \quad \quad  \left[ \begin{array}{l} \frac{n}{\bs{2}} \cdot
\Bmlcaqr{\frac{m}{\sqrt{\sProc{2}}}}{\bs{2}}{\bs{1}}{\Proc{1}}\\
\quad + \sum_{r=2}^{l} \frac{n}{\bs{r}} \cdot \lBRecBCast{T}{1\ldots r}{\frac{m \bs{r}}{ \sqrt{\sProc{r}}}} + \frac{n}{\bs{2}} \cdot
\Bupfact{\frac{m}{\sqrt{\sProc{2}}}}{\frac{n}{\sqrt{\sProc{2}}}}{\bs{1}}{\Proc{1}} \end{array}
\right] \\
\quad \quad \quad +  \sum_{r=2}^{l-1} \frac{n \cdot N_r}{\bs{r}} \left[ \begin{array}{l}
          \lBRecComm{T}{1\ldots r}{\frac{\bs{r}^2}{2}} + \frac{\bs{r}}{\bs{2}} \cdot
  \Bmlcaqr{\frac{2\bs{r}}{\prod_{j=2}^{r-1}\sqrt{\Proc{j}}}}{\bs{2}}{\bs{1}}{\Proc{1}} \\
          \quad +  \lBRecBCast{T}{1\ldots r}{\bs{r}^2} + \frac{\bs{r}}{\bs{2}} \cdot
\Bupfact{\frac{\bs{r}}{\prod_{j=2}^{r-1}\sqrt{\Proc{j}}}}{\frac{n}{\sqrt{\sProc{2}}}}{\bs{1}}{\Proc{1}}\\
          \quad \quad + \lBRecComm{T}{1\ldots r}{\frac{\bs{r} n}{\sqrt{\sProc{r}}}} + \frac{\bs{r}}{\bs{2}} \cdot
\Bupelim{\frac{2\bs{r}}{\prod_{j=2}^{r-1}\sqrt{\Proc{j}}}}{\frac{n}{\sqrt{\sProc{2}}}}{\bs{1}}{\Proc{1}}
\end{array} \right] \\
\quad \quad \quad \quad + \frac{n \cdot N_l}{\bs{l}} \left[ \begin{array}{l} \lBRecComm{T}{1\ldots r}{\frac{\bs{l}^2}{2}} + \frac{\bs{l}}{\bs{2}} \cdot
 \Bmlcaqr{\frac{2\bs{l}}{\prod_{j=2}^{l-1}\sqrt{\Proc{j}}}}{\bs{2}}{\bs{1}}{\Proc{1}} \\
\quad + \lBRecComm{T}{1\ldots r}{\frac{\bs{l} n}{\sqrt{\Proc{l}}}}
+ \frac{\bs{l}}{\bs{2}} \cdot
\Bupelim{\frac{2\bs{l}}{\prod_{j=2}^{l-1}\sqrt{\Proc{j}}}}{\frac{n}{\sqrt{\sProc{2}}}}{\bs{1}}{\Proc{1}}
\end{array} \right]
\end{array}
\end{equation*}

\end{appendix}

\end{document}